# Two-Dimensional Semiconducting Metal Organic Frameworks with Auxetic Effect, Room Temperature Ferrimagnetism, Chiral Ferroelectricity, Bipolar Spin Polarization and Topological Nodal Lines/Points

Xiangyang Li,[†,∆] Qing-Bo Liu,[&,∆] Yongsen Tang,[∥,∆] Wei Li,[‡] Ning Ding,[¶] Zhao Liu,[†] Hua-Hua Fu,[&] Shuai Dong,[¶] Xingxing Li,[*,†,§,∥] and Jinlong Yang[*,†,§,∥]

[†]Hefei National Research Center for Physical Sciences at the Microscale, University of Science and Technology of China, Hefei, Anhui 230026, China
[∥]Hefei National Laboratory, University of Science and Technology of China, Hefei 230088, China
[§]Department of Chemical Physics, University of Science and Technology of China, Hefei, Anhui 230026, China
[‡]Department of Physics, University of Science and Technology of China, Hefei, Anhui 230026, China
[&]School of Physics and Wuhan National High Magnetic Field Center, Huazhong University of Science and Technology, Wuhan 430074, China
[∥]Laboratory of Solid-State Microstructures and Innovative Center of Advanced Microstructures, Nanjing University, Nanjing, 210093, China
[¶]School of Physics, Southeast University, Nanjing 211189, China

**ABSTRACT:** Two-dimensional (2D) semiconductors integrated with two or more functions are the cornerstone for constructing multifunctional nanodevices, but remain largely limited. Here, by tuning the spin state of organic linkers and the symmetry/topology of crystal lattice, we predict a class of unprecedented multifunctional semiconductors in 2D Cr(II) five-membered heterocyclic metal organic frameworks that simultaneously possess auxetic effect, room temperature ferrimagnetism, chiral ferroelectricity, electrically reversible spin polarization and topological nodal lines/points. Taking 2D Cr(TDZ)$_2$ (TDZ=1.2.5-thiadiazole) as an exemplification, the auxetic effect is produced by the anti-tetra-chiral lattice structure. The high temperature ferrimagnetism originates from the strong $d$-$p$ direct magnetic exchange interaction between Cr cations and TDZ doublet radical anions. Meanwhile, the clockwise-counterclockwise alignment of TDZ' dipoles results in unique 2D chiral ferroelectricity with atomic-scale vortex-antivortex states. 2D Cr(TDZ)$_2$ is an intrinsic bipolar magnetic semiconductor where half-metallic conduction with switchable spin-polarization direction can be induced by applying a gate voltage. Besides, the symmetry of the little group $C_4$ of lattice structure endows 2D Cr(TDZ)$_2$ with topological nodal lines and a quadratic nodal point in the Brillouin zone near the Fermi level.

## INTRODUCTION

Two-dimensional (2D) multifunctional materials with unique atomic-scale configurations and exotic electronic properties have aroused great interests in recent decades.[1,2] However, up to now, only limited numbers of such materials have been reported experimentally or theoretically, such as NiI$_2$,[3] ReWCl$_6$,[4] $h$-Ti$_2$(O$_2$)$_3$,[5] and AlB$_6$.[6] In addition, most are concentrated in traditional inorganic compounds with only two or three functions (Table S1). Developing 2D multifunctional materials with more functions and exotic properties remains a pending task.

Considering the structural rigidity and limited tunability of inorganic compounds, we turn our attention to organometallic materials with structural variability and rich functionalization possibilities.[7,8] Organometallic frameworks are hybrid porous materials composed of abundant metal nodes and inexpensive organic linkers.[8] By tuning metal nodes or organic linkers or the connectivity between them,[9-15] they can possess functional properties with potential applications in fields of emergent electromechanical, magnetoelectronic, magnetic sensing, and topological quantum technologies. For instance, by using the dicyanoquinonediimine as a rotatory unit, the Cr(dicyanoquinonediimine)$_2$ sheet has been predicted to be an auxetic magnet.[9] By changing the spin state of organic linkers from singlet to doublet and introducing a strong $d$-$p$ direct ferrimagnetic exchange interaction, high Curie temperature ($T_C$) magnetic semiconductors Cr(pentalene)$_2$ ($T_C$=560 K),[10] Cr(diketopyrrolopyrrole)$_2$ ($T_C$=316 K)[11] and Cr(pyrazine)$_2$ ($T_C$=342 K)[12,13] have been designed theoretically. Via distorting out-of-plane K$^+$ counterions, ferroelectric 2D magnetic K$_3$M$_2$[PcMO$_8$] (M = Cr-Co) sheets have been forecasted.[14] Through forming a Kagome lattice on a superconducting substrate, the experimentally synthesized 2D Cu$_2$(dicyanoanthracene)$_3$ sheet has been calculated and found to possess topological Dirac cones coupled with substrate's superconductivity.[16]

Among numerous organometallic frameworks, transition metal Cr atom as a common metal node has been widely used,[9-14] and its related planar tetracoordinate molecules or crystals have been extensively synthesized.[17-19] Particularly, Perlepe *et al.* have prepared a layered Li$_{0.7}$[Cr(pyrazine)$_2$]Cl$_{0.7}$·0.25(THF) (THF=tetrahydrofuran) crystal with room temperature ferrimagnetism, in which each Cr(II) is coordinated to four pyrazine organic linkers within the layers forming a planar tetracoordinated sheet.[19] If the inversion symmetric pyrazine rings are replaced by inversion symmetry-breaking organic linkers to increase the tunable degree of freedom of the crystal structure, the functional properties can be further enriched.

In this work, by employing the Cr(II) as nodes and inversion symmetry-breaking five-membered aromatic heterocycles (1.2.5-thiadiazole, 1,2,5-oxadiazole, 1,2,5-selenadiazole) as organic linkers, a class of unprecedented 2D semiconductors with up to five important functions, i.e. auxetic effect, room-temperature ferrimagnetism, chiral ferroelectricity, electrical field controlled spin polarization, and topological nodal lines/points are predicted in metal organic frameworks with a planar tetracoordinate structure. As exemplified by $Cr(TDZ)_2$ (TDZ=1.2.5-thiadiazole) sheet, due to the anti-tetra-chiral square lattice, auxetic effect emerges along the diagonal direction with a negative Poisson's ratio (NPR) of about -0.12. At the same time, the strong $d$-$p$ direct magnetic exchange interaction between Cr cations and TDZ doublet radicals enables a room-temperature ferrimagnetism with $T_C$ = 378 K. Moreover, 2D chiral ferroelectricity with atomic-scale vortex-antivortex states is discovered as a result of the coexistence of clockwise-counterclockwise dipoles, which has previously been shown to exist only in extremely rare cases and complex heterojunctions.[20, 21] The electronic band structure indicates 2D $Cr(TDZ)_2$ not only belongs to a special class of bipolar magnetic semiconductors (BMSs)[22, 23] with the carriers' spin orientation readily reversible by electrical gating, but also be a topological material with square nodal lines and a quadratic nodal point protected by $C_4$ crystal symmetry in the first Brillouin zone near the Fermi level. For practical applications, these multifunctional materials provide an excellent platform to study the proximity effect between different properties. Moreover, by combining different functions, some high-performance spintronic devices can be designed, such as ultra-high-density data storage device.

COMPUTATIONAL METHODS

Density functional theory (DFT) calculations are performed by using the projector-augmented wave method and the Perdew-Burke-Ernzerhof (PBE) functional as implemented in Vienna *ab initio* Simulation Package (VASP).[24, 25] The approach of Grimme (DFT-D3) with Becke-Jonson damping is adopted for the van der Waals (vdW) interactions.[26] To treat the partially filled 3$d$ orbitals of transition metal atoms, the strongly correlated correction is considered with PBE+U method.[27] The values of effective exchange interaction parameter ($J$) and onsite Coulomb interaction parameter ($U$) are respectively set as 1.0 and 3.0 eV, which are the same as previously calculated $Cr(pyz)_2$ sheet.[12, 13] The energy cutoff for plane-wave basis set is 520 eV. For the first Brillouin zone integration, the Monkhorst-Pack $k$-point mesh is used with a grid spacing less than 0.02 Å$^{-1}$. The energy and force criteria for convergence are set to $1\times10^{-6}$ eV and 0.01 eV/Å, respectively. A vacuum region of about 15 Å is chosen to avoid mirror interactions between periodic layers. The phonon spectrum is simulated by using the finite displacement method as implemented in Phonopy package interfaced with VASP.[28] A 2×2×1 supercell with a Monkhorst-Pack $k$-point mesh of 2×2×1 is adopted. The thermal stability is assessed according to the *ab initio* molecular dynamic (AIMD) simulation at 300 K by using a 3×3×1 supercell. The climbing image nudged elastic band (CI-NEB) method is adopted to investigate the structure of transition state and the energy barrier between different ferroelectric phases.[29] Considering the complicated transition between different ferroelectric phases, a series of possible structures on the intermediate path are conjectured to find the structure with the highest energy, and their phonon spectra are further analyzed to verify that they are indeed saddle points (Figure S1). The out of plane electric polarization is evaluated by using the dipole correction scheme.[30] To accurately calculate the electronic structure, the screened hybrid HSE06 functional is applied,[31] which includes the accurate Fock exchange and usually performs much better than the PBE and PBE+U methods.[32-34] The edge states have been performed using the open-source code WANNIERTOOLS[35] based on the Wannier tight-binding model constructed using the WANNIER90 code.[36] The irreps of the electronic bands are computed by the program IR2TB on the electronic Hamiltonian of the tight-binding model.[37]

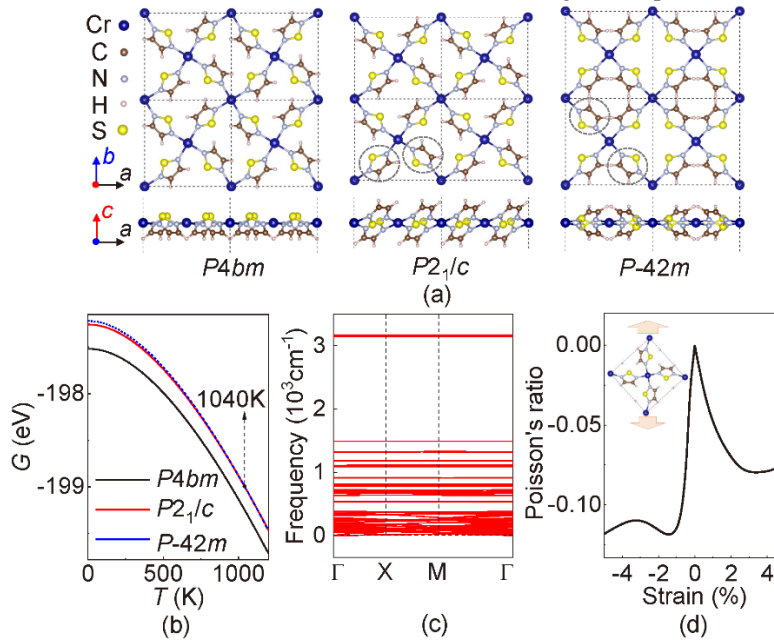

**Figure 1.** (a) Geometrical structures of the three phases of $Cr(TDZ)_2$ (TDZ = 1.2.5-thiadiazole). (b) Gibbs free energies per unit cell of the three phases of $Cr(TDZ)_2$ as a function of temperature. The dotted arrow marks the temperature of the phase transition point of *P*-42*m* to *P*2$_1$/*c*. (c) Phonon spectrum of the most stable *P*4*bm* phase. (d) Poisson's ratio of the *P*4*bm* structure over a ±5% strain range along the diagonal direction.

RESULTS AND DISCUSSION

Due to the lack of inversion symmetry, each five-membered aromatic heterocyclic structure has two different orientations with respect to the lattice plane, leading to the diversification of crystal structures formed with Cr atoms. Taking TDZ organic ring as an example, Figure 1(a) shows three low energy structures of the Cr(TDZ)$_2$ sheet with point groups of *P4bm*, *P2$_1$/c*, and *P-42m* symmetry, respectively. In these structures, four TDZ organic rings form approximately square planar coordination with the Cr atoms, and each TDZ unit is connected by two adjacent Cr atoms. The four organic rings are arranged clockwise or counterclockwise around the Cr atoms, with the ring planes presenting an inclination angle of about 43° with respect to the *ab* lattice plane. The unit cell parameters are $a=b=9.14$ Å for the *P4bm* phase, $a$ ($b$)=9.20 (8.66) Å for the *P2$_1$/c* phase, and $a=b=9.24$ Å for the *P-42m* phase. For the *P4bm* structure, the S atoms in TDZ rings are all on one side of the *ab* plane. The panel on the left side of Figure 1(a) shows one possible structure in which all S atoms are located on the upper side of the lattice plane. Such spatial inversion symmetry-breaking gives rise to the electric polarization and ferroelectricity in the structure (to be elaborated later). In contrast, when the S atoms in two ortho or para TDZ rings are on the other side of lattice plane [see the rings enclosed by dotted circles in the middle and right panels of Figure 1(a)], the corresponding *P2$_1$/c* and *P-42m* structures are formed and tend to be antiferroelectric.

First-principles calculations identify that the Gibbs free energy of *P4bm* crystal at 0 K is 0.26 and 0.30 eV per unit cell lower than those of *P2$_1$/c* and *P-42m* crystals, respectively. As the temperature increases, the *P4bm* structure remains to possess the lowest energy [Figure 1(b)], thus it is the ground state configuration. For the other two structures, a phase transition from *P2$_1$/c* to *P-42m* is observed at around 1040 K. In the following studies, we mainly focus on the most stable *P4bm* phase.

As shown in Figure 1(c), no obvious imaginary frequency is observed from the calculated phonon spectrum, indicating that Cr(TDZ)$_2$ sheet is dynamically stable. Due to the rather large lattice constant, the phonon bands are dispersionless. The existence of a large number of soft phonon modes illustrates the flexibility of Cr(TDZ)$_2$ sheet.[10] For example, the maximum Young's modulus of *P4bm* structure is only 39 GPa (Figure S2), which is much smaller than that of MoS$_2$ (170-370 GPa).[38] Besides the dynamic stability, the thermal stability is further examined by performing AIMD simulation at 300 K (Figure S3). It is found that during the simulation, the total energy always fluctuates near its equilibrium value without a sudden drop, and the lattice structure can maintain well without any reconstruction after 9 *ps*, confirming that the structure is thermally stable.

Interestingly, the structure of Cr(TDZ)$_2$ sheet belongs to the so called anti-tetra-chiral lattice capable of exhibiting auxeticity,[39] therefore a distinct auxetic effect and negative Poisson's ratio (NPR) along the diagonal direction is expected [see Figure 1(d)]. Here, Poisson's ratio is defined as $\partial\varepsilon_t/\partial\varepsilon_a$, where $\varepsilon_t$ and $\varepsilon_a$ are strains in the transverse and corresponding longitudinal directions, respectively. The maximum value of NPR can reach -0.12 in the 5% strain range. This value is smaller than that of 2D Cr(dicyanoquinonediimine)$_2$ (-0.85),[9] but comparable to most reported 2D inorganic auxetic materials, such as Ag$_2$S (-0.12),[40] Be$_5$C$_2$ (-0.16),[41] and SnSe (-0.17).[42] The auxetic property endows 2D Cr(TDZ)$_2$ with potential applications in nano-mechanics, defense and aerospace aspects.

In Cr(TDZ)$_2$ sheet, each Cr atom donates formally two electrons to neighboring TDZ rings, forming a Cr(II) cation and TDZ doublet radical anions. The total magnetic moment of each chemical formula is 2 $\mu_B$, in which the magnetic moment of the Cr(II) cation is 3.4 $\mu_B$ and that of each TDZ radicals is -0.7 $\mu_B$. Therefore, the spins of Cr and TDZ can be approximated as 2 and 1/2, respectively. To determine the magnetic ground state of Cr(TDZ)$_2$ sheet, five different magnetic states, including one ferromagnetic (FM) state, one antiferromagnetic state, and three ferrimagnetic (FiM) states (Figure S4), are investigated. The results show that the FiM1 state is the ground state, where the spins on Cr(II) cations are all antiparallelly aligned with the spins on TDZ radicals. Figure 2(a) gives the spin density distribution of the FiM1 state. Obviously, the spin density on the TDZ radicals is distributed over the $p_z$ orbitals of all nonmetallic C, N and S atoms. Due to the strong direct exchange interaction between the π-conjugated orbitals of TDZ rings and *d* orbitals of Cr, the FiM1 state is significantly more stable than the FM state by as large as 0.85 eV per chemical formula.

For practical spintronic applications, it is crucial to keep the magnetic ordering of Cr(TDZ)$_2$ sheet above room temperature. To confirm this, we perform Monte Carlo simulations

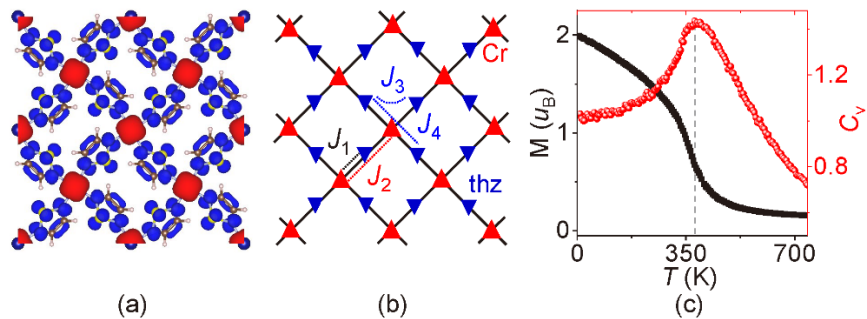

**Figure 2.** (a) Spin density distribution of Cr(TDZ)$_2$ sheet with a *P4bm* symmetry in the ground ferrimagnetic state. Red and blue indicate up and down spins, respectively. (b) The nearest neighbor and next-nearest neighbor spin exchange paths for Cr(TDZ)$_2$ sheet. The exchange-coupling parameters $J_k$ ($k = 1\sim4$) are also marked. $J_1$ represents the interaction between the Cr atom and nearest neighbor TDZ. $J_2$ means the interaction between the nearest two Cr atoms. $J_3$ and $J_4$ serve as the interaction between the nearest and next-nearest two TDZ. (c) Magnetic moment ($M$) per unit cell (black) and specific heat $C_v$ (red) as a function of temperature by using a Monte Carlo simulation based on the classic Heisenberg model. The magnetic exchange parameters used here are calculated with the HSE06 functional.

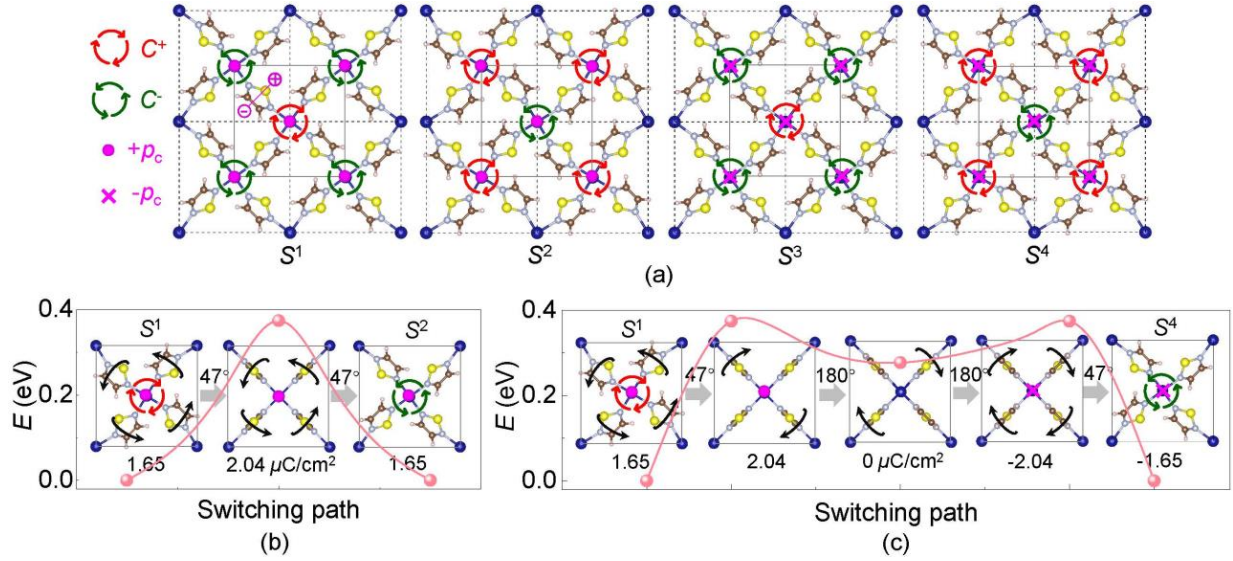

**Figure 3.** (a) Four possible chiral vortex-antivortex states $S^i$ ($i = 1\sim4$) of Cr(TDZ)$_2$ sheet with a *P4bm* symmetry. C$^+$ represents a vortex state comprising four clockwise dipoles around the central Cr atoms and C$^-$ means an anti-vortex state comprising four counterclockwise dipoles around the central Cr atoms. $+p_c$ ($-p_c$) serves as a state of the total electric polarization that is outward (inward) perpendicular to the *ab* lattice plane. (b) A possible path and energy barrier for the transition from $S^1$ to $S^2$. (c) A possible path and energy barrier for the transition from $S^1$ to $S^4$.

based on the classic Heisenberg model Hamiltonian,[43, 44]

$$H = -\sum_k \sum_{i>j} \sum_j J_k \times S_i \cdot S_j + \sum_i D_i S_{iz}^2 \quad (1)$$

where $J_k$ are four different exchange-coupling parameters presented in Figure 2(b), and $S_i$ is the spin of Cr or TDZ. $D_i$ are magnetic anisotropy parameters with a value of 123.5 $\mu$eV for Cr and 0 $\mu$eV for TDZ, as derived from the calculated magnetic anisotropy energy of 494.1 $\mu$eV per formula. The values of $J_k$ are deduced from the energy differences between the above mentioned five different magnetic states (summarized in Table S2). Figure 2(c) shows the temperature-dependent spin magnetic moment (*M*) per chemical formula, in which the *M* gradually decreases from 2 $\mu_B$ to 0 with the increase of temperature. The specific heat $C_v = (\langle E^2 \rangle - \langle E \rangle^2)/T^2$ is obtained after the system reaching equilibrium at a given temperature, and the peak position manifests that the ferrimagnetic-paramagnetic transition occurs at the Curie temperature $T_C$ of 378 K, significantly higher than room temperature.

More intriguingly, due to the inversion symmetry-breaking feature of the five membered heterocycles, the negative charge center of the TDZ ring does not coincide with the positive one, thus an intrinsic electric polarization would be present. Therefore, each TDZ ring can be regarded as an electric dipole, which is inclined from the c-axis at an angle of 47° and reversible in dipole direction by rotating the TDZ rings. The superimposed c-axis component of all TDZ dipoles induces the ferroelectric polarization ($\pm p_c$) of Cr(TDZ)$_2$ sheet along the *c*-direction. In the *ab* plane, four non-collinear electric dipoles connecting one of the Cr atoms form an atomic-scale vortex state $C^+$, and another four electric dipoles linking the adjacent Cr atoms form an antivortex state $C^-$ [Figure 3(a)]. Such two states assemble a 2D lattice with chiral vortex-antivortex polar states down to the monolayer scale, which is distinct from the nanometer-scale chiral vortex-antivortex arrays in a complex heterojunction structure with alternating lead titanate and strontium titanate layers.[21]

Based on the above analysis, the non-collinear ferroelectric states of Cr(TDZ)$_2$ sheet can be described by two ferroelectric order parameters $Q_1 = +C^++C^-$ in the *ab* plane and $Q_2 = +p_c+p_c$ along the *c*-direction, where $Q_1$ and $Q_2$ obviously have two degenerate modes $\pm Q_1$ and $\pm Q_2$, respectively. As shown in Figure 3(a), those four order parameters can describe four degenerate ferroelectric vortex states: $S^1$ (+$Q_1$, +$Q_2$), $S^2$ (+$Q_1$, -$Q_2$), $S^3$ (-$Q_1$, +$Q_2$), $S^4$ (-$Q_1$, -$Q_2$). It is worth mentioning that the in-depth study of such ferroelectric materials with chiral vortex properties at the atomic scale is of great value for the understanding of non-collinear ferroelectric and chiral ferroelectric physics.

DFT calculations indicate that the polarization values of the above four vortex states are all 1.65 $\mu$C/cm$^2$, where the polarization directions of $S^1$ and $S^2$ are along the positive direction of *c*-axis, while those of $S^3$ and $S^4$ are along the negative direction of *c*-axis. The polarization values are comparable to those of reported Sc$_2$CO$_2$ (1.6 $\mu$C/cm$^2$),[45] ReWCl$_6$ (3.22 $\mu$C/cm$^2$),[4] and hexagonal YMnO$_3$ (5 $\mu$C/cm$^2$).[46]

Figure 3(b) shows the transition path from $S^1$ to $S^2$. The four TDZ rings in the $S^1$ state first simultaneously rotate 47° counterclockwise to reach the transition state and then further rotate 47° counterclockwise to evolve to the $S^2$ state, where the energy barrier is 0.37 eV per TDZ. At the transition state, all the TDZ rings are perpendicular to the *ab* plane, where the polarization reaches the maximum value of 2.04 $\mu$C/cm$^2$ and the vortex state feature disappears. The transition from $S^1$ to $S^4$ is complicated, and the corresponding energy barrier is also 0.37 eV per TDZ. Figure 3(c) illustrates one possible path. First the four TDZ rings in the $S^1$ state rotate 47° counterclockwise, next the two TDZ rings at the para position rotate 180° clockwise, then the other two TDZ rings rotate 180° clockwise, and finally the four TDZ rings simultaneously rotate 47° clockwise to get to the $S^4$ state. Specifically, a series of possible intermediate states exist throughout the transition, but none of them is stable after examining their phonon spectra. For example, an intermediate state without polarization [see the middle panel in Figure 3(c)] possesses an energy of 0.27 eV higher than the ground structure, and its phonon spectrum presents a certain imaginary frequency; see Figure S5. In addition, Figure S6 displays another possible transition path, where the energy barrier remains to be 0.37 eV. Thus, we can infer that the energy barriers from $S^i$ to $S^j$ are all around 0.37 eV, the value of which is higher than that of WO$_2$Cl$_2$ (0.22 eV),[47] but lower than those of K$_3$Fe$_2$[PcFeO$_8$] (0.38 eV)[14] and CrI$_3$ (0.65 eV).[48]

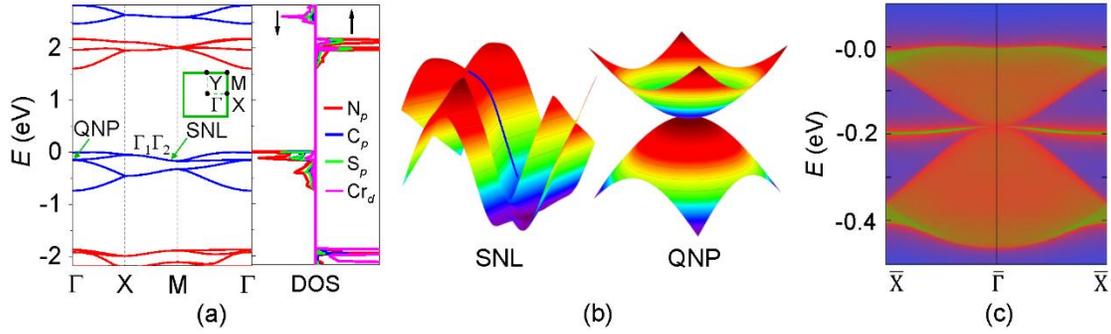

**Figure 4.** (a) Spin-polarized band structures and projected density of states for Cr(TDZ)$_2$ sheet with the HSE06 functional. Red and black lines represent spin-up and spin-down bands, respectively. The inset shows high-symmetry points in the first Brillouin zone. QNP and SNL stand for quadratic nodal point and square nodal line, respectively. (b) Three-dimensional energy band structures of an SNL and a QNP near the Fermi level. (c) Dirac-cone edge states for QNP.

To reveal the electronic properties of Cr(TDZ)$_2$ sheet, the band structures and density of states are calculated by using the HSE06 functional, as shown in Figure 4(a). Obviously, it is a direct semiconductor with a band gap of 1.60 eV. The valence and conduction bands are 100% spin polarized in opposite spin channels, suggesting the Cr(TDZ)$_2$ sheet belongs to an intrinsic bipolar magnetic semiconductor (BMS),[22, 23] where electrical gating can generate half-metallic conduction with controllable spin polarization directions. Moreover, the energy bands near the Fermi level not only possess doubly-degenerate nodal lines on the X-M path protected by a 2D irreps $\Gamma_1\Gamma_2$ in little group $C_4$, but also exist a quadratic nodal point (QNP) with a zero Chern number protected by a 2D irreps $\Gamma_5$ in point group $C_{4v}$ at the $\Gamma$ point (Note S1). The nodal lines are distributed at the boundary of 2D Brillouin zone, showing a square-shaped feature; see the inset of Figure 4(a). By investigating the distributions of projected density of states, one can derive that the states of square nodal lines (SNLs) and QNP are dominated by the $p$ orbitals of N, C, and S atoms. For clarity, Figure 4(b) demonstrates the 3D band characteristics of an SNL and a QNP near the Fermi level.

To further verify the topological properties, we calculate the edge states of Cr(TDZ)$_2$ sheet along the (100) direction. As displayed in Figure 4(c), the QNP at the $\bar{\Gamma}$ point forms a clear quadratic Dirac cone edge state, which proves that it is nontrivial. Since the SNLs are projected into the one-dimensional Brillouin zone, the corresponding edge states cannot be observed. Besides, after considering the spin-orbit coupling (SOC) effect, the QNP of Cr(TDZ)$_2$ opens a topological gap of 7 meV (Figure S7), which is comparable to those of Mn(C$_6$H$_5$)$_3$ (9.5 meV),[49] Mn$_2$C$_6$S$_{12}$ (7~15 meV),[50] and Cr$_2$Se$_3$ (6.7 meV).[51]

Based on the example of Cr(TDZ)$_2$ sheet, we can easily expand into a range of multifunctional organometallic semiconductors by substituting the TDZ organic linkers with other five-membered heterocycles, such as Cr(ODZ)$_2$ (ODZ=1,2,5-oxadiazole) and Cr(SDZ)$_2$ (SDZ=1,2,5-selenadiazole). For the Cr(ODZ)$_2$ sheet, the ground state is the antiferroelectric $P2_1/c$ state, which is 0.26 eV lower in energy than the ferroelectric $P4bm$ state. In contrast, the ground state of Cr(SDZ)$_2$ sheet is ferroelectric with an electric polarization strength of about 1.01 $\mu$C/cm$^2$ (Table S3), slightly weaker than that of Cr(TDZ)$_2$ sheet. The transition energy barriers between different ferroelectric phases remain around 0.37 eV. The auxetic effect of these two extended sheets is superior to that of Cr(TDZ)$_2$ sheet, where the maximum absolute value of NPR reaches 0.17 and 0.13 for Cr(ODZ)$_2$ and Cr(SDZ)$_2$ sheets, respectively (Figure S8). Both $T_C$ are above room temperature with the highest being 410 K (Figure S9). Meanwhile, the Cr(ODZ)$_2$ and Cr(SDZ)$_2$ sheets are also BMSs with oppositely spin-polarized valence and conduction band edges (Figure S10). Besides, compared with Cr(TDZ)$_2$, the QNP of Cr(SDZ)$_2$ opens a larger topological gap of 33 meV from the HSE+SOC band structure, providing a potential platform for studying the anomalous quantum Hall effect.

For the experimental fabrication of these 2D organometallic frameworks, one possible route is to adopt top-down technologies. Similar to the synthesized Li$_{0.7}$[Cr(pyz)$_2$]Cl$_{0.7}$·0.25(THF) (THF=tetrahydrofuran) crystal,[19] their bulk layered crystals are firstly realized by combining redox-active coordination chemistry[52] and postsynthetic reduction modification,[19] and then the corresponding sheets are achieved through the mechanical exfoliation. Another possible route is to use bottom-up methodologies. Metal atoms and organic linker molecules are deposited onto a metal surface by molecular beam evaporation or electron beam evaporation to induce their self-assembly to form 2D organometallic frameworks. Such preparation strategies have been widely used to synthesize similar coordination structures, e.g., Mn-TCNQ$_4$ network,[53] Ni-TPyP network,[54] TPA-Cs, BDA-Cs, and TDA-Cs networks.[55]

In practical applications, integrating so many functional properties into a single sheet can offer at least two advantages. The first is to provide an ideal platform to study different kinds of proximity effect.[56] Specifically, as shown in Figure 5(a), the proximity effects between FiM, ferroelectricity, chirality, BMS, and topology can be investigated by constructing a bilayer homojunction. Such a homojunction can effectively avoid additional effects caused by lattice mismatch of the het-erojunction. The second is to improve the performance of related spintronic devices through the synergy between multiple functions. For instance, when the five functional properties of FiM, ferroelectricity, chirality, BMS, and topology are simultaneously applied to a data storage device, the storage density can be increased 16 (2$^4$) folds compared to a single-function device since each function contains two switchable states. The switching between different storage states can be realized under an external electric or magnetic field. Figure 5(b) displays a field effect transistor with the Cr(TDZ)$_2$ sheet as a channel material to illustrate the specific modulation method in different functions. The orientation of spin moment ($\pm M_s$) can be changed by applying a magnetic field $\vec{B}$ perpendicular to the sheet plane. The transition between different electrical polarization ($\pm P$) states and chiral ($C^{\pm}$) states can be achieved by assigning a certain electric field. Both the direction of current spin polarization ($I_{\uparrow/\downarrow}$) and the transition from trivial semiconductor (SC) to topological half-metal including boundary states can be modulated by using different gate voltages.

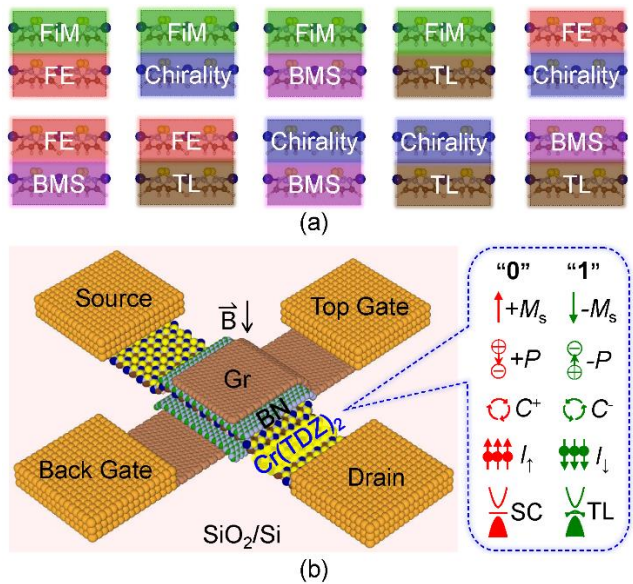

**Figure 5.** (a) Constructing a Cr(TDZ)$_2$-based homojunction to explore the proximity effect among FiM, ferroelectricity (FE), chirality, BMS, and topology (TL). (b) 3D schematic diagram of an field-effect transistor device based on the Cr(TDZ)$_2$ sheet. The field-effect transistor is fabricated by sequentially vertically stacking graphene (Gr), BN, Cr(TDZ)$_2$, BN, and Gr sheets on a SiO$_2$/Si substrate. The top and back gates are connected to one of the Gr sheets, respectively. The source and drain electrodes are linked to the Cr(TDZ)$_2$ sheet. The inset on the right shows five pairs of possible storage parameters, that is, spin magnetic moment ($\pm M_s$), electric polarization ($\pm P$), chirality ($C^{\pm}$), spin polarized current ($I_{\uparrow/\downarrow}$), and semiconductor (SC)/topological half-metal.

CONCLUSIONS

To summarize, on the basis of first-principles calculations, we report a class of unprecedented 2D multifunctional semiconductors with several unique properties including auxetic effect, room temperature ferrimagnetism, chiral ferroelectricity, electrically controllable spin polarization and topological nodal lines/points. The simultaneous realization of these functions relies on the combined tuning of the spin state of organic linkers and the symmetry/topology of lattice structure in metal organic frameworks constructed by Cr(II) and inversion symmetry-breaking five-membered aromatic heterocycles (1,2,5-thiadiazole, 1,2,5-oxadiazole, 1,2,5-selenadiazole). These materials not only serve as promising candidates for studying different proximity effects and designing multifunctional nanodevices, but also imply the unique abilities of metal organic frameworks in obtaining electronic/magnetic properties that are difficult to achieve in inorganic materials such as chiral vortex-antivortex polar states in the monolayer limit.

## ASSOCIATED CONTENT

**Supporting Information**. Additional material includes phonon spectra of the transition state and an intermediate state, Young's modulus, AIMD simulation, and HSE+SOC band structure for Cr(TDZ)$_2$; spin density of different magnetic states; another possible pathway for transition from $S^1$ to $S^4$; Poisson's ratio, Curie temperature, HSE and HSE+SOC band structures for Cr(ODZ)$_2$ and Cr(SDZ)$_2$; magnetic exchange parameters and ground state properties of different sheets; two-band $k·p$ Hamiltonian for SNLs and QNP. This material is available free of charge via the Internet at http://pubs.acs.org.

## AUTHOR INFORMATION

Corresponding Author

* lixx@ustc.edu.cn
* jlyang@ustc.edu.cn
$^\Delta$ These authors contributed equally to this work.

Notes

The authors declare no competing financial interest.

## ACKNOWLEDGMENT

This work is supported by Anhui Initiative in Quantum Information Technologies with Grant No. AHY090400, by the Youth Innovation Promotion Association CAS with Grant No. 2019441, by the Innovation Program for Quantum Science and Technology with Grant No. 2021ZD0303306, by USTC Research Funds of the Double First-Class Initiative with Grant No. YD2060002011, by the National Natural Science Foundation of China with Grant No. 12147113, and by project funded by the China Postdoctoral Science Foundation with Grant No. 2021M691149. The computational resources are provided by the Supercomputing Center of University of Science and Technology of China, Supercomputing Center of Chinese Academy of Sciences, and Tianjin and Shanghai Supercomputer Centers.

Table of Contents artwork

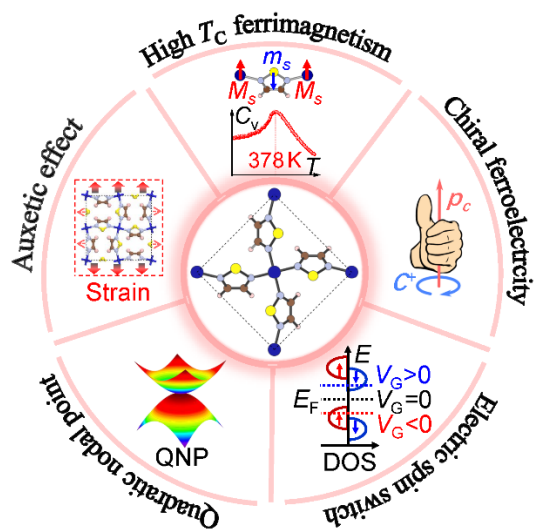